\documentclass[12pt]{book}

\usepackage[dvips]{graphicx,color}
\usepackage{makeidx,hori,zon}

\makeauthorindex

\BookTitle{Inflating Horizon of Particle Astrophysics and Cosmology}
\CopyRight{\copyright 2005 by Universal Academy Press, Inc.\ and Yamada Science Foundation}

\begin{document}

\pagenumbering{arabic}

\chapter{
Primordial Black Holes: Do They Exist and Are They Useful?}

\author{%
Bernard CARR\\
{\it Astronomy Unit, Queen Mary, University of London,  Mile End Road, London E1 4NS, England\\
Research Center for the Early Universe, University of Tokyo, Tokyo 113-0033, Japan\\
B.J.Carr@qmul.ac.uk}}
%
%
\AuthorContents{B.J.Carr} 

\AuthorIndex{Carr}{B.J.} 

\section*{Abstract}

Recent developments in the study of primordial black holes (PBHs) are
reviewed, with particular emphasis on their formation and evaporation. It is still not clear whether PBHs
formed but, if they did, they could provide a unique probe of the early Universe, gravitational collapse, high energy physics and quantum gravity. Indeed their study may place
interesting constraints on the physics relevant to these areas even if they
never existed.

\section*{Preface}

It is a great honour to speak at this meeting because Katsuhiko Sato has been a friend and colleague for 25 years. I first met him when he visited Cambridge in 1980. At the time he was a guest professor at NORDITA and I was a research fellow at the Institute of Astronomy. It was a particularly important period for Katsu because he had just written his famous paper on inflation (arguably the first paper on the topic, although its publication was delayed). I remember our first meeting very vividly: he came to my office and asked to take my photograph because he said I was famous. Nobody had ever told me I was famous before, so I was quite chuffed! 
Of course, Katsu is now much more famous than me, so I intend to repay the compliment by taking a photograph of him at this meeting!

Although we have not written any papers together, Katsu was the first Japanese physicist with whom I interacted and he has had an important influence on my career. This is because meeting him was one of the factors which prompted me to spend a year at Kyoto University in 1982 and I have made many subsequent visits to Japan. On several of these visits I have enjoyed the hospitality of Katsu and his wife, so I would like to express my gratitude for this. Indeed I would say that Katsu's scientific contribution lies not only in his own work but also in the role he has played in encouraging collaboration with Western scientists (many of whom are attending this conference).  In my case, this collaboration is particularly strong, since two years ago I married a Japanese lady. It is strange to reflect that this might not have happened without the chain of events triggered by Katsu coming to my office in 1980! 

I've chosen to talk about primordial black holes, partly because this is one of my own lifelong interests, but also because it is an area in which Japanese physicists have made a vital contribution. Katsu has himself written many papers on the subject \cite{ms,sskm,kssm,kss,mssk,is,sns,send} and I will highlight some of these in this paper. PBHs also bear on inflation, the topic of the meeting: on the one hand, the fluctuations which arise at inflation are the most likely source of PBH; on the other hand, upper limits on the number of PBHs place constraints on inflationary models. 

\section{Introduction}

Black holes with a wide range of masses could have formed in the early Universe as a result of the great compression associated with the Big Bang \cite{h71,zn}. A 
comparison of the cosmological density at a time $t$ after the Big Bang with the density associated with a black hole of mass $M$
shows that PBHs would have of order the particle horizon mass at their formation epoch:
\begin{equation}
M_H(t) \approx {c^3 t\over G} \approx 10^{15}\left({t\over 10^{-
23} \ {\rm s}}\right) g.
\end{equation}
PBHs could thus span an enormous mass range: those formed at the
Planck time ($10^{-43}$s) would have the Planck mass ($10^{-5}$g),
whereas those formed at 1~s would be as large as $10^5M_{\odot}$,
comparable to the mass of the holes thought to reside in galactic nuclei. 
By contrast, black holes forming at the present epoch could never be smaller than about $1M_{\odot}$.

The realization that PBHs might be small prompted Hawking
to study their quantum properties. This led to his famous 
discovery \cite{h74}
that black holes radiate thermally with a temperature
\begin{equation}
T = {\hbar c^3\over 8\pi GMk}
\approx 10^{-7} \left({M\over M_{\odot}}\right)^{-1} {\rm K},
\end{equation}
so they evaporate on a timescale 
\begin{equation}
      \tau(M) \approx {\hbar c^4\over G^2M^3}
\approx 10^{64} \left({M\over M_{\odot}}\right)^3~\mbox{y}.
 \end{equation}
Only black holes smaller
than about $10^{15}$g would have evaporated by the present epoch, so eqn (1) implies
that this effect could be important only for black holes which formed before $10^{-23}$s.

Hawking's result was a tremendous conceptual advance, since it linked three previously disparate areas of physics - quantum theory, general relativity and thermodynamics. Even if PBHs never existed, it has been useful to think about them! However, at first sight it was bad news for PBH enthusiasts. For since PBHs with a mass of $10^{15}$g
would be producing photons with energy of order 100~MeV at the present epoch, the observational limit on
the $\gamma$-ray background intensity at 100~MeV immediately 
implied
that their density could not exceed $10^{-8}$ times the
critical density \cite{ph}. Not only did this render 
PBHs
unlikely dark matter candidates, it also implied that there was 
little
chance of detecting black hole explosions at the present epoch \cite{pw}.
Nevertheless, it was soon realized that the $\gamma$-ray background results did not preclude PBHs playing other important cosmological roles \cite{c76} and some of these will be discussed here. 

\section{How PBHs Form}

The high density of the early Universe is a necessary but not sufficient condition for PBH formation. One also needs density fluctuations, so that overdense regions can eventually stop expanding and recollapse. Indeed one reason for studying PBH formation and evaporation is that it imposes important constraints on primordial inhomogeneities. PBHs may also form at various phase transitions expected to occur in the early Universe. In some of these one require pre-existing density fluctuations, but in others the PBHs form spontaneously, even if the Universe starts off perfectly smooth. In the latter case, limits on the number of PBHs constrain any parameters associated with the phase transition. The formation of PBHs from inhomogeneities is discussed in Section 4, but we now briefly review the other mechanisms.

{\it Soft equation of state.}
Some phase transitions can lead to the equation of state becoming soft for a while. For example, the pressure may be reduced if the Universe's mass is ever channelled into particles which are massive enough to be non-relativistic. In such cases, the effect of pressure in stopping collapse is unimportant and the probability of PBH formation just depends upon the fraction of regions which are sufficiently spherical to undergo collapse \cite{kp}. For a given spectrum of primordial fluctuations, this means that there may just be a narrow mass range - associated with the period of the soft equation of state - in which the PBHs form.

{\it Collapse of cosmic loops.}
 In the cosmic string scenario, one expects some strings to self-intersect and form cosmic loops. A typical loop will be larger than its Schwarzschild radius by the factor $(G\mu)^{-1}$, where $\mu$ is the string mass per unit length. If strings play a role in generating large-scale structure, $G\mu$  must be of order $10^{-6}$. However, as discussed by many authors \cite{pz,h89,gs,cc,mbw}, there is always a small probability that a cosmic loop will get into a configuration in which every dimension lies within its Schwarzschild radius. This probability depends upon both $\mu$ and the string correlation scale. Note that the holes form with equal probability at every epoch, so they should have an extended mass spectrum.

{\it Bubble collisions.}  
Bubbles of broken symmetry might arise at any spontaneously broken symmetry epoch and various people have suggested that PBHs could form as a result of bubble collisions \cite{cs,hms,kss,ls,m94}. However, this happens only if the bubble formation rate per Hubble volume is finely tuned: if it is much larger than the Hubble rate, the entire Universe undergoes the phase transition immediately and there is not time to form black holes; if it is much less than the Hubble rate, the bubbles are very rare and never collide. The holes should have a mass of order the horizon mass at the phase transition, so PBHs forming at the GUT epoch would have a mass of $10^3$g, those forming at the electroweak unification epoch would have a mass of $10^{28}$g, and those forming at the QCD (quark-hadron) phase transition would have mass of around $1M_{\odot}$. There could also be wormhole production at a 1st order phase transition \cite{kssm,mssk}.  Recently the production of PBHs from bubble collisons at the end of 1st order inflation has been studied extensively by Khlopov and his colleagues \cite{kon,kkrs}. 

{\it Collapse of domain walls}
The collapse of sufficiently large closed domain walls produced at a 2nd order phase transition in the vacuum state of a scalar field, such as might be associated with inflation, could lead to PBH formation \cite{der,rks}. These PBHs would have a small mass for a thermal phase transition with the usual equilibrium conditions. However, they could be much larger if one invoked a non-equlibrium scenario \cite{rsk}. Indeed Khlopov et al. argue that they could span a wide range of masses, with a fractal structure of smaller PBHs clustered around larger ones \cite{krs}. 

\section{Why PBHs are Useful}

The study of PBHs provides a unique probe of four areas of physics: the early Universe; quantum gravity; gravitational collapse and high energy physics. One can probe the last two topics only if PBHs exist today but one can gain insight into the first two topics even if PBHs never formed.  

{\it PBHs as a probe of the early Universe ($M<10^{15}g$).}
Many processes in the early Universe could be modified by PBH evaporations \cite{npsz}. For example, recent studies have considered how evaporations could change the details of baryosynthesis \cite{dol} and
nucleosynthesis \cite{ms,ky}, provide a source of gravitinos \cite{kb} and neutrinos  \cite{bk}, swallow monopoles \cite{is,sf} and remove domain walls by puncturing them \cite{stoy}. PBHs evaporating at later times could have important astrophysical effects, such as helping to reionize the universe \cite{hf}. There has also been interest in whether PBHs could preserve gravitational ``memory" if the value of $G$ was different in the early Universes \cite{b92,bc,cg,jac,hcg}. 
Other constraints are associated with thermodynamics \cite{lee} and the holographic principle \cite{cuho}.

{\it PBHs as a probe of gravitational collapse ($M>10^{15}g$).}
This relates to recent developments in the study of ``critical phenomena" and the issue of whether PBHs are viable dark matter candidates. In the latter case, their gravitational effects could lead to distinctive dynamical, lensing and gravitational-wave signatures. They could also influence the development of large-scale structure and the formation of supermassive black holes in galactic nuclei. These issues are discussed in Section 5. 

{\it PBHS as a probe of high energy physics ($M\sim 10^{15}g$).}
PBH evaporating today could contribute to cosmic rays, whose energy distribution would then give significant information about the high energy physics involved in the final explosive phase of black hole evaporation. In particular, PBHs could contribute to the cosmological and Galactic $\gamma$-ray backgrounds, the antiprotons and positrons in cosmic rays, gamma-ray bursts, and the annihilation-line radiation coming from centre of the Galaxy. These issues are discussed in Section 6.

{\it PBHS as a  probe of quantum gravity ($M\sim10^{-5}g$).}
Many new factors could come into play when a black hole's mass gets down to the Planck regime, including the effects of extra dimensions and quantum-gravitational spacetime fluctuations. For example, it has been suggested that black hole evaporation could cease at this point, in which case Planck relics could contribute to the dark matter. More radically, it is possible that quantum gravity effects could appear at the TeV scale and this leads to the intriguing possibility that small black holes could be generated in accelerators experiments or cosmic ray events. Although such black holes are not technically ``primordial", this would have radical implications for PBHs themselves.  These issues are discussed in Section 7.

\section{Formation of PBHs from Inhomogeneities}

One of the most important reasons for studying PBHs is that it enables one to 
place limits on the spectrum of density fluctuations in the early Universe. This
is because, if the PBHs form
directly from density perturbations, the fraction of
regions undergoing collapse at any epoch is determined 
by
the root-mean-square amplitude $\epsilon$ of the fluctuations
entering the horizon at that epoch and the equation of state 
$p=\gamma \rho~(0< \gamma <1)$. One usually expects a radiation equation of
state ($\gamma =1/3$) in the early Universe but it may have deviated from this in some periods. 

\subsection{Simplistic Analysis}

Early calculations assumed that the overdense region which evolves to a PBH is spherically symmetric and part of a closed Friedmann model. In order to collapse
against the pressure, such a region must be larger than the
Jeans length at maximum expansion and this is just 
$\sqrt{\gamma}$
times the horizon size. On the other hand, it cannot be larger than the horizon size, else it would
form a separate closed universe and not be part of our Universe \cite{ch}. 

This has two important implications. First, PBHs forming at time $t$ after the Big Bang should
have of order the horizon mass given by eqn (1).
Second, for a region destined to collapse to a PBH, one requires the fractional overdensity at the horizon
epoch $\delta$ to exceed $\gamma$. Providing the density 
fluctuations have a Gaussian distribution and are spherically
symmetric, one can infer that the fraction of regions of mass 
$M$ which collapse is \cite{c75}
\begin{equation}
\beta(M) \sim \epsilon(M) \exp\left[-{\gamma^2\over 
2\epsilon(M)^2}\right]
\end{equation}
where $\epsilon (M)$ is the value of $\epsilon$ when the 
horizon
mass is $M$. The PBHs can have an extended mass spectrum only 
if the fluctuations are scale-invariant (i.e. with $\epsilon$ 
independent of $M$). 
In this case, the PBH mass distribution is given by \cite{c75}
\begin{equation}
dn/dM = (\alpha -2) (M/M_*)^{-\alpha}M_*^{-2}\Omega_{\rm PBH}\rho_{{\rm crit}}
\end{equation}
where $M_* \approx 10^{15}$g is the current lower cut-off in the mass spectrum due to evaporations, $\Omega_{PBH}$ is the total density of the PBHs in units of the critical density (which depends on $\beta$) and the exponent $\alpha$ is  determined by the equation of state:
\begin{equation}
\alpha = \left({1+3\gamma \over 1+\gamma}\right) + 1 .
\end{equation}
$\alpha=5/2$ if one has a radiation equation of state. This means that the density of PBHs larger than $M$ falls off as $M^{-1/2}$, so most of the PBH density is contained in the smallest ones. 

Many scenarios for the cosmological density fluctuations predict that $\epsilon$ is at least approximately scale-invariant but the sensitive dependence of $\beta$ on $\epsilon$ means that even tiny deviations from scale-invariance can be important.  If $\epsilon (M)$ decreases with increasing $M$, then the spectrum falls off exponentially and most of the PBH density is contained in the smallest ones. If $\epsilon (M)$ increases with increasing $M$, the spectrum rises exponentially and - if PBHs were to form at all - they could only do so at large scales. However, the microwave background anisotropies would then be larger than observed, so this is unlikely. It is also possible to enhance PBH formation at particular mass-scales by introducing bumps in $\epsilon (M)$.

\begin{figure}[htbp]
\begin{center}
\includegraphics[scale=0.5]{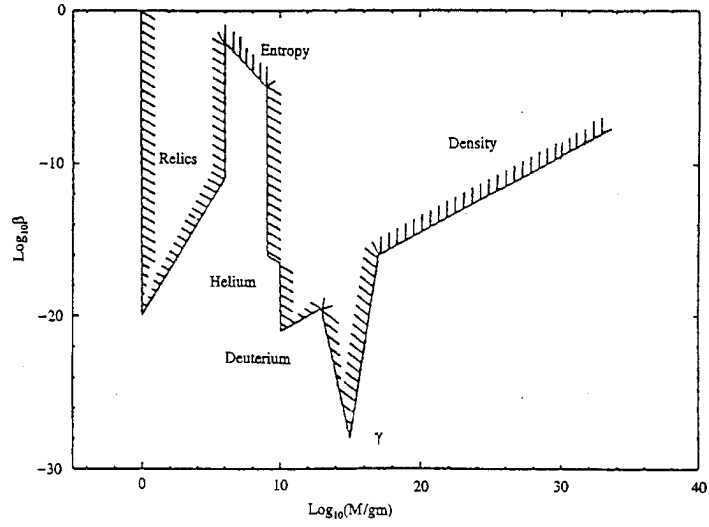}
\caption{Constraints on $\beta(M)$}
\label{fig:bc1}
\end{center}
\end{figure}

The current density parameter $\Omega_{\rm PBH}$ associated
with PBHs which form at a redshift $z$ or time $t$ is related to
$\beta$ by \cite{c75}
\begin{equation}
\Omega_{\rm PBH} = \beta\Omega_R(1+z) \approx 10^6 
\beta\left({t\over s}\right)^{-1/2} \approx 10^{18}\beta\left({M\over 10^{15}g}\right)^{-1/2}
\end{equation}
where $\Omega_R \approx 10^{-4}$ is the density parameter of the microwave
background and we have used eqn (1). The $(1+z)$ factor arises 
because the radiation density scales as $(1+z)^4$, whereas the PBH
density scales as $(1+z)^3$. Any limit on $\Omega_{\rm PBH}$ therefore places a constraint on $\beta(M)$ and 
the constraints are summarized in Fig.~\ref{fig:bc1}, which is taken from Carr et al. \cite{cgl}.
The constraint for non-evaporating mass ranges above
$10^{15}$g comes from requiring
$\Omega_{\rm PBH}<1$ but stronger constraints are associated with PBHs
smaller than this since they would have evaporated by now. 
The strongest one is the $\gamma$-ray limit associated with the $10^{15}$g PBHs evaporating
at the present epoch \cite{ph}. Other ones are associated with the generation of entropy and modifications to the cosmological production of light 
elements \cite{npsz}. The
constraints below $10^{6}$g are based on the (uncertain) assumption that evaporating PBHs leave stable Planck
mass relics, an issue which is discussed in detail in Section 7.1.

\begin{figure}[htbp]
\begin{center}
\includegraphics[scale=0.5]{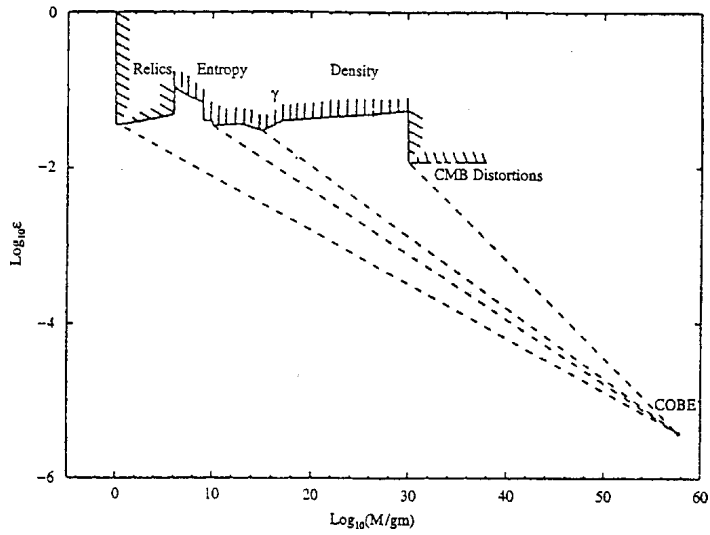}
\caption{Constraints on $\epsilon(M)$}
\label{fig:bc2}
\end{center}
\end{figure}

The constraints on 
$\beta(M)$ can be converted into constraints on 
$\epsilon(M)$ using eqn (4) and these are shown in Fig.~\ref{fig:bc2} Also shown here are the (non-PBH) constraints
associated with the spectral distortions in the cosmic microwave background induced by the dissipation of 
intermediate scale density perturbations and the COBE quadrupole measurement. This shows that one needs the
fluctuation amplitude to decrease with increasing scale in order to produce PBHs and the lines corresponding
to various slopes in the $\epsilon (M)$ relationship are also shown in Fig.~\ref{fig:bc2}

\subsection{Refinements of Simplistic Analysis}

The criterion for PBH formation given above is rather simplistic and needs to be tested with detailed numerical calculations. The first hydrodynamical studies of PBH formation were carried out by Nadezhin et al. \cite{nnp}. These roughly confirmed the criterion $\delta > \gamma$ for PBH formation, although the PBHs could be be somewhat smaller than the horizon. In recent years several groups have carried out more detailed hydrodynamical calculations and these have refined the $\delta > \gamma$ criterion, suggesting that one needs $\delta >0.7$ rather than $\delta >0.3$ \cite{nj99,ss}. This in turn affects the estimate for $\beta(M)$ given by eqn (4).

A particularly interesting development has been the application of ``critical phenomena"
to PBH formation. Studies of the collapse of various types of spherically symmetric matter fields have shown that there
is  always a critical solution which separates those configurations which form a black hole from 
those which disperse to an asymptotically flat state. The configurations are described by some index $p$ and, as the
critical index $p_c$ is approached, the black hole mass is found to scale as $(p-p_c)^{\eta}$ for some exponent
$\eta$. This effect was first discovered for scalar fields \cite{chop} but subsequently
demonstrated for radiation \cite{ec} and then more general fluids with equation of state $p=\gamma \rho$ \cite{mai,kha}. 

In all these studies the spacetime was assumed to be asymptotically flat. However, Niemeyer \& Jedamzik \cite{nj98} have applied
the same idea to study black hole formation in asymptotically Friedmann models and have found similar results. 
For a variety of initial density perturbation profiles, they find that the
relationship between the PBH mass and the the horizon-scale density perturbation has the form
\begin{equation}
M = K M_H(\delta - \delta_c)^{\eta}
\end{equation}
where $M_H$ is the horizon mass and the constants are in the range $0.34<\eta<0.37$, $2.4<K<11.9$ and
$0.67<\delta_c <0.71$ for the various configurations. More recently, Musco et al. \cite{mmr} have found that the critical overdensity lies in the lower range $0.43<\delta_c <0.47$ if one only allows growing modes at decoupling (which is more plausible if the fluctuations derive from inflation). They also find that the exponent $\eta$ is modified if there is a cosmological constant. Since $M \rightarrow 0$ as $\delta \rightarrow \delta_c$, 
the existence of critical phenomena suggests that PBHs may be much smaller than the particle horizon at formation and this also modifies the mass spectrum \cite{yok,kir,gl99,g00}. Although Hawke \& Stewart \cite{hs} claim that the formation of shocks prevents black holes forming on scales below $10^{-4}$ of the horizon mass, this has been disputed \cite{mmr}. 

It should be stressed that the description of fluctuations beyond the horizon is somewhat problematic and it is clearer to use a gauge-invariant description which involves the total energy or metric perturbation \cite{ss}. Also the derivation of the mass spectrum given by eqn (4) is based on Press-Schechter theory and it is more satisfactory to use peaks theory. Both these points have been considered by Green et al. \cite{glms}. They find that that the critical value for the density contrast is around 0.3, which is close to the value originally advocated 30 years ago!

Another refinement of the simplistic analysis which underlies eqn (4) concerns the assumption that the fluctuations have a Gaussian distribution. Bullock \& Primack \cite{bp2} and Ivanov \cite{i98} have pointed out that this may not apply if the fluctuations derive from an inflationary period (discussed in Section 4.3). So long as the fluctuations are small ($\delta \phi /\phi \ll1$), as certainly applies on a galactic scale, this assumption is valid. However, for PBH formation one requires  $\delta \phi /\phi \sim 1$, and, in this case, the coupling of different Fourier modes destroys the Gaussianity. Their analysis suggests that $\beta(M)$ can be very different from the value indicated by eqn (4) but it still depends very sensitively on $\epsilon$.

\subsection{PBHs and Inflation}

Inflation has two important consequences for PBHs. On the one hand, any PBHs formed before the end of inflation will be diluted to a negligible density. Inflation thus imposes a lower limit on the PBH mass spectrum:
\begin{equation}
M > M_{\rm{min}} = M_{P}(T_{RH}/T_{P})^{-2}			 		
\end{equation}
where $T_{RH}$ is the reheat temperature and $T_{P}\approx 10^{19}$ GeV is the Planck temperature. The CMB quadrupole measurement implies $T_{RH}\approx 10^{16}$GeV, so 
$M_{\rm{min}}$ certainly exceeds $1$ g. On the other hand, inflation will itself generate fluctuations and these may suffice to produce PBHs after reheating. If the inflaton potential is $V(\phi)$, then the horizon-scale fluctuations for a mass-scale $M$ are
\begin{equation}
\epsilon(M) \approx \left(\frac{V^{3/2}}{M_{P}^3 V'}\right)_H
\end{equation}						
where a prime denotes $d/d \phi$  and the right-hand side is evaluated for the value of $\phi$ when the mass-scale $M$ falls within the horizon. 
In the standard chaotic inflationary scenario, one makes the ``slow-roll" and ``friction-dominated" assumptions:
\begin{equation}
\xi \equiv (M_{P}V'/V)^2 \ll 1,  \quad    \eta \equiv M_{P}^2 V''/V \ll 1	.	
\end{equation}
Usually the exponent $n$ characterizing the power spectrum of the fluctuations, 
$|\delta _k|^2 \approx k^n$, is very close to but slightly below 1:
\begin{equation}
n = 1 + 4\xi - 2\eta \approx 1.							
\end{equation}
Since $\epsilon$ scales as $M^{(1-n)/4}$, this means that the fluctuations are slightly increasing with scale. The normalization required to explain galaxy formation ($\epsilon \approx 10^{-5}$) would then preclude the formation of PBHs on a smaller scale. 
If PBH formation is to occur, one needs the fluctuations to decrease with increasing mass ($n>1$) and, from eqn (12), this is only possible if the scalar field is accelerating sufficiently fast that \cite{cl}
\begin{equation}
V''/V > (1/2) (V'/V)^2  .							
\end{equation}
This condition is certainly satisfied in some scenarios \cite{gil} and, if it is, eqn (4) implies that the PBH density will be dominated by the ones forming immediately after reheating. This is because the volume dilution of the PBHs forming shortly before the end of inflation will dominate the enhancement associated with eqn (4). However, it should be stressed that the validity of eqn (10) at the very end of inflation is questionable since the usual assumptions may fail then \cite{kolb,lmsz}. 

\begin{figure}[htbp]
\begin{center}
\includegraphics[scale=0.6]{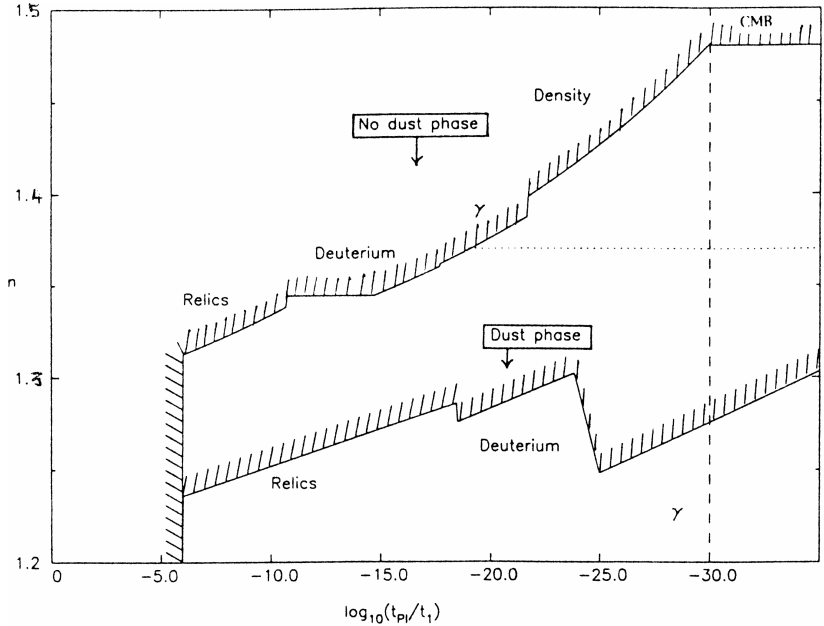}
\caption{Constraints on spectral index $n$ in terms of reheat time $t_1$}
\label{fig:bc3}
\end{center}
\end{figure}

Since each value of $n$ corresponds to a straight line in Fig.\ref{fig:bc3}, any particular value for the reheat time $t_1$ corresponds to an upper limit on $n$.
This limit is indicated in Fig.\ref{fig:bc3}, which is taken from \cite{cgl}, apart from a correction pointed out by Green \& Liddle \cite{gl97}. Similar constraints have been obtained by several other people \cite{klm,bkp}. The figure also shows how the constraint on $n$ is strengthened if the reheating at the end of inflation is sufficiently slow for there to be a dust-like phase \cite{cgl,glr}.
However, it should be stressed that not all inflationary scenarios predict that the spectral index should be constant. Indeed, Hodges \& Blumenthal \cite{hb} have pointed out that one can get any spectrum for the fluctuations by suitably choosing the form of $V(\phi)$. For example, eqn (10) suggests that one can get a spike in the spectrum by flattening the potential over some mass range (since the fluctuation diverges when $V'$ goes to 0). This idea was exploited by Ivanov et al. \cite{inn}, who fine-tuned the position of the spike so that it corresponds to the mass-scale associated with microlensing events observed in the Large Magellanic Cloud \cite{alc}. 

It should be noted that the relationship between the variance of the mass fluctuations relevant for PBHs and the present day horizon-scale density fluctuations in the inflation scenario is not trivial. For scale-free fluctuations, there is a simple relationship between them but it is more complicated otherwise. 
Blais et al. \cite{bbkp} find that the mass fluctuations are reduced by 34\% for a spectral index in the range $1<n<1.3$, while Polarski \cite{pol} finds a further 15\% reduction if there is a  cosmological constant. 

Even if PBHs never actually formed as a result of inflation, studying them places important constraints on the many types of inflationary scenarios. Besides the chaotic scenario discussed above, there are also the variants of inflation described as designer \cite{hb,inn,y99}, supernatural \cite{rsg}, supersymmetric \cite{g99}, hybrid \cite{glw,kky}, multiple \cite{y98}, oscillating \cite{tar}, ghost \cite{muk}, running mass \cite{lgl} and saddle \cite{eks}. There are also scenarios in which the inflaton serves as dark matter \cite{lmu}. PBH formation has been studied in all of these models. Note that in the standard scenario inflation ends by the decay of the inflaton into radiation. However, in the preheating scenario inflation ends more rapidly because of resonant coupling between the inflaton and another scalar field. This generates extra fluctuations, which are not of the form indicated by eqn (10) and these might also produce PBHs  \cite{ep,bt,fk,gm}. However, a recent analysis by Suyama et al. \cite{stbk}, incorporating a full non-linear lattice simulation for various models, suggests that PBH formation is insignificant. There have also been a variety of other proposals for generating perturbations at the end of inflation \cite{krv,lyth,salem} and the probability of PBH formation would then again be unrelated to eqn (10). 

\section{PBHs and Dark Matter}

Roughly $30\%$ of the total density of the Universe is now thought to be in the form of ``cold dark matter''. Recently there has been a lot of interest in whether PBHs could provide this, since those larger than $10^{15}$g would not have evaporated yet and would certainly be massive enough to be dynamically ``cold''.  It is also possible that the Planck relics of evaporated PBHs could provide the dark matter and this is discussed in Section 7.1.

\subsection{Constraints on PBHs in the Galactic Halo}
There are some mass ranges in which PBHs are already excluded from providing dark matter in galactic halos. For example, femtolensing of gamma-ray bursts by PBHs precludes those in the mass range $10^{17}-10^{20}$g from having a critical density and microlensing of stars in the Large Magellanic Cloud, while allowing a tenth-critical-density at around a solar mass,  excludes $10^{26}-10^{34}$g PBHs \cite{alc}. However, there are no constraints in the intermediate (sublunar) mass range $10^{20}-10^{26}$g \cite{bkp}. Note that LISA might detect the gravitational impulse induced by any nearby passing PBH \cite{ab,sc}. However, this method would not work below $10^{14}$g (because the effect would be hidden by the Moon) or above $10^{20}$g (because the encounters would be too rare). 

One possibility is that  PBHs with a mass of around $1M_{\odot}$ could have 
formed at the quark-hadron phase transition at $10^{-5}$s because of a temporary softening of the equation of state then.
If the QCD phase 
transition is assumed to be 1st order, then hydrodynamical calculations show that the value of
$\delta$ required for PBH formation is reduced below the value which pertains in the radiation era \cite{jn99}. This means that 
PBH formation will be strongly enhanced at the QCD epoch, with the mass distribution peaking at around the horizon mass then \cite{j97,y97,pf} . Such PBHs might be able to explain the MACHO microlensing results \cite{alc} and also the claimed microlensing of quasars \cite{haw}. Another possibility is that PBHs with a mass of around $10^{-7}M_{\odot}$ could form in TeV quantum gravity scenarios \cite{it}. 

One of the interesting
implications of these scenarios is the possible existence of a halo population of {\it binary} black
holes \cite{nak}. With a full halo of such objects, there could be a huge number of binaries inside 50 kpc and some of these could be coalescing due to
gravitational radiation losses at the present epoch \cite{bc2}. If the associated gravitational waves were detected, it would provide a unique probe of the halo distribution \cite{itn}. Gravity waves from binary PBHs would be detectable down to $10^{-5}M_{\odot}$ using VIRGO, $10^{-7}M_{\odot}$ using EURO and $10^{-11}M_{\odot}$ using LISA \cite{it}. 

\subsection{Large-Scale Structure and Supermassive Black Holes}

PBHs might also play a role in the formation of large-scale structure \cite{dol,rsk,kr,krs}. For example, it has been realized for some time that Poisson fluctuations in the number density of PBHs can generate appreciable density perturbations on large scales if the PBHs are big enough \cite{mes,c77,fre,csi}. This idea has recently been invoked to explain voids and Lyman-alpha clouds \cite{ams}.  In such scenarios, it is important to know how much a PBH can grow through accretion and it has been argued that this could be significant even in the period after matter-radiation equality \cite{mo}.

Several people have suggested that the $10^6 - 10^{8}M_{\odot}$ supermassive black holes thought to reside in galactic nuclei could be of primordial origin. For example, a 1st order phase transition could produce clusters of PBHs, around which a single supermassive black hole might then condense \cite{der}. Alternatively, it has been proposed that inflation could produce closed domain walls, which then collapse to form $10^8M_{\odot}$ PBHs \cite{krs}). Another scheme invokes more modest mass PBHs of $10^3 M_{\odot}$, resulting from a feature in the inflaton potential, which then grow to $10^{8}M_{\odot}$ by accretion \cite{duc}. A related idea comes from Bean and Magueijo, who have suggested \cite{bm} that PBHs may accrete from the quintessence field which is invoked to explain the acceleration of the Universe. However, they use a Newtonian formula for the accretion rate \cite{zn} and this is known to be questionable \cite{ch}. Recent studies of the accretion of a scalar field by a PBH also indicate that this is unlikely \cite{hc,hmc}. 
 
\section{Cosmic Rays from PBHS}

A black hole of mass $M$ will emit particles
like a black-body of temperature \cite{h75}
\begin{equation}
T \approx 10^{26} \left({M\over g}\right)^{-1}\ {\rm K}
\approx \left({M\over 10^{13} g}\right)^{-1} \ {\rm 
GeV}.
\end{equation}
This assumes that the hole has no charge or angular
momentum. This is a reasonable assumption since charge and 
angular
momentum will also be lost through quantum emission but on a 
shorter
timescale than the mass \cite{p77}. This means that it loses mass at a rate
\begin{equation}
      \dot{M} = -5\times 10^{25}(M/g)^{-2}f(M)~\mbox{g~s}^{-1}
\end{equation}
where the factor $f(M)$ depends on the number of particle 
species which are light enough to be emitted by a hole of mass 
$M$, so the 
lifetime is
\begin{equation}
      \tau(M) = 6 \times 10^{-27} f (M)^{-1}(M/g)^3~\mbox{s}.
\end{equation}
The factor $f$ is normalized to be 1 for holes larger than 
$10^{17}$~g
and such holes are only able to emit ``massless" particles like
photons, neutrinos and gravitons. Holes in the mass range
$10^{15}~\mbox{g} < M < 10^{17}~\mbox{g}$ are also able to emit
electrons, while those in the range $10^{14}~\mbox{g} < M <
10^{15}~\mbox{g}$ emit muons which subsequently decay into 
electrons
and neutrinos. The latter range includes, in particular, the critical
mass $M_*$ for which $\tau$ equals the age of the Universe. If one assumes the currently favoured age of 13.7 Gyr, this corresponds to $M_{*} = 5 \times 10^{14}$g \cite{mcp}.  

Once $M$ falls below $10^{14}$g, a black hole can also begin to emit
hadrons. However, hadrons are composite particles made up of 
quarks
held together by gluons. For temperatures exceeding the QCD 
confinement scale of $\Lambda_{\rm QCD} = 250-300$~GeV, one 
would
therefore expect these fundamental particles to be emitted rather 
than
composite particles.  Only pions would be light enough to be 
emitted
below $\Lambda_{\rm QCD}$. Since there are 12 quark degrees of 
freedom per
flavour and 16 gluon degrees of freedom, one would also expect 
the
emission rate (i.e. the value of $f$) to increase dramatically once 
the
QCD temperature is reached.
One can
regard the black hole as emitting quark and gluon jets of the kind
produced in collider events \cite{m91}.  The jets will decay into hadrons over 
a distance which is always much larger than the size of the hole, so gravitational 
effects can be neglected. The hadrons may then decay into astrophysically stable particles through weak and electomagnetic decays.

To find the final spectra of stable particles emitted from a black
hole, one must convolve the Hawking emission spectrum with the jet fragmentation function. This gives the instantaneous
emission spectrum shown in Fig.4 for a $T=1$~GeV black 
hole \cite{mw}. The direct emission just 
corresponds to
the small bumps on the right. All the particle spectra show a peak at
100~MeV due to pion decays; the electrons and neutrinos also 
have peaks at 1~MeV due to neutron decays.
In order to determine the present day background spectrum of
particles generated by PBH evaporations, one must first 
integrate over
the lifetime of each hole of mass $M$ and then over the PBH 
mass
spectrum \cite{mw}. In doing so, one must allow for the 
fact
that smaller holes will evaporate at an earlier cosmological epoch, 
so
the particles they generate will be redshifted in energy by the
present epoch. 

\begin{figure}[htbp]
\begin{center}
\includegraphics[scale=0.5]{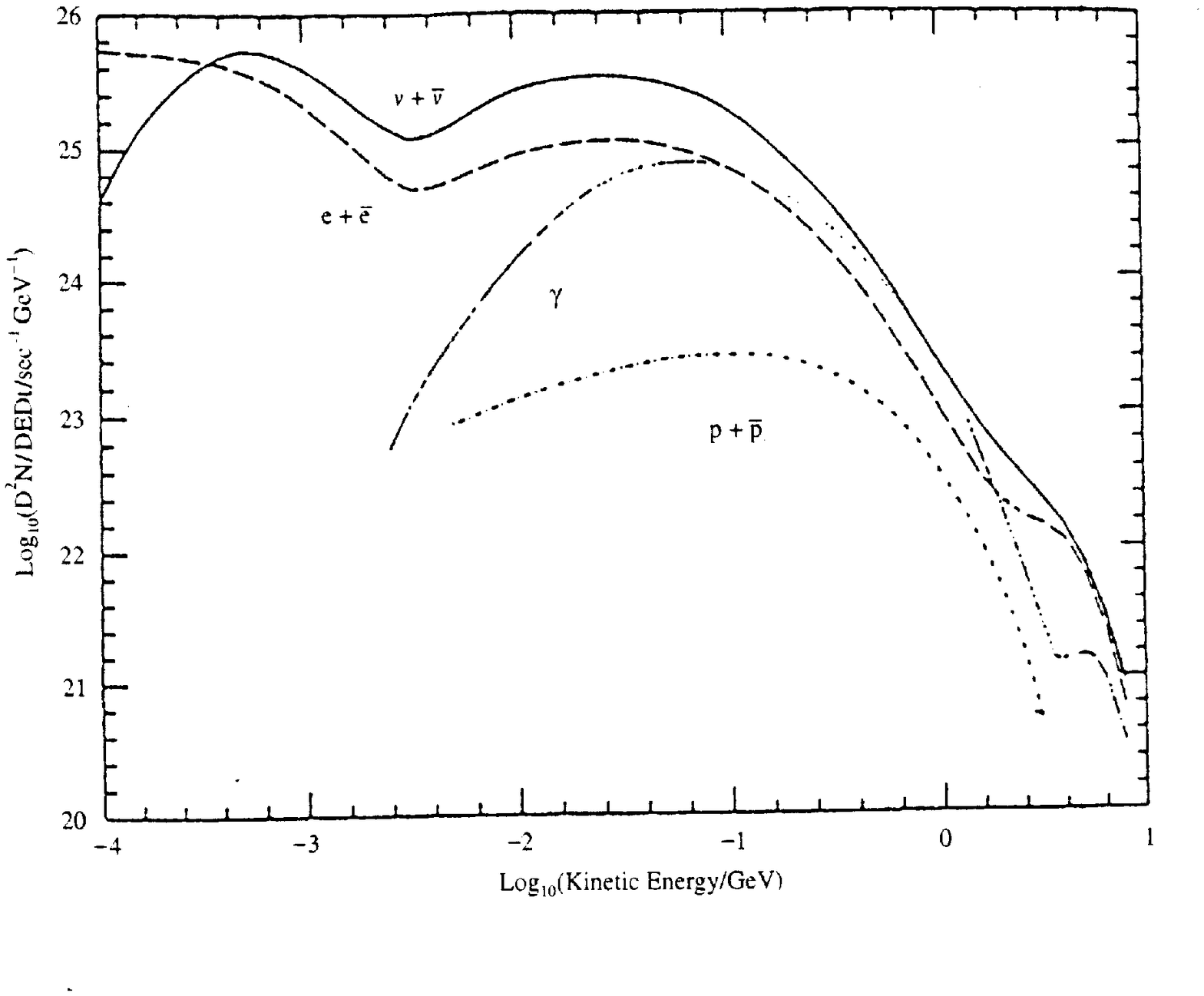}
\caption{Instantaneous emission from a 1~GeV black hole}
\label{fig:bc5}
\end{center}
\end{figure}

\begin{figure}[htbp]
\begin{center}
\includegraphics[scale=0.5]{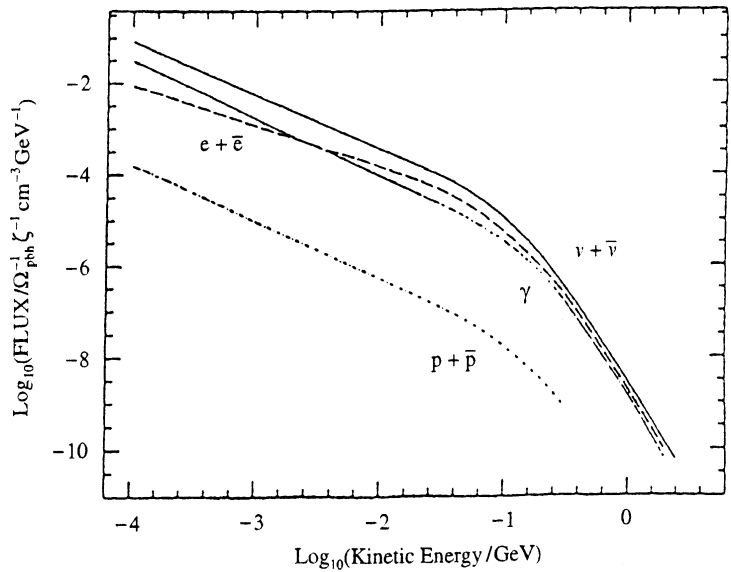}
\caption{Spectrum of particles from uniformly distributed PBHs}
\label{fig:bc6}
\end{center}
\end{figure}

If the holes are uniformly distributed throughout the
Universe, the background spectra should have the form indicated 
in Fig.\ref{fig:bc6} All the spectra have rather similar shapes: an $E^{-3}$
fall-off for $E > 100$~MeV due to the final phases of evaporation 
at the present epoch and an $E^{-1}$ tail for $E<100$~MeV due to the
fragmentation of jets produced at the present and earlier epochs. 
Note
that the $E^{-1}$ tail generally masks any effect associated with the mass spectrum of smaller PBHs which evaporated at earlier epochs \cite{c76}.
The situation is more complicated if the PBHs evaporating at 
the present epoch are clustered inside our own Galactic halo (as is 
most
likely). In this case, any charged particles emitted after the epoch
of galaxy formation (i.e. from PBHs only somewhat smaller than $M_*$) will have their flux enhanced relative to the
photon spectra by a factor $\xi$ which depends upon the halo
concentration factor and the time for which particles are trapped
inside the halo by the Galactic magnetic field.
This time is rather uncertain and also energy-dependent. At 100~MeV
one has $\xi \sim 10^3$ for electrons or positrons 
and $\xi \sim
10^4$ for protons and antiprotons.  MacGibbon \& Carr \cite{mc} first used the observed cosmic ray spectra to constrain $\Omega_{\rm PBH}$ but their estimates have been subsequently refined.

\subsection{Gamma-Rays}
EGRET observations \cite{sre} of the $\gamma$-ray background between 30 MeV and 120 GeV leads to an upper limit $\Omega_{\rm PBH} \le (5.1 \pm1.3) \times 10^{-9} h^{-2}$, 
where $h$ is the Hubble parameter in units of 100 \cite{cm}. This is a refinement of the original Page-Hawking limit, but the form of the spectrum suggests that PBHs do not
provide the dominant contribution. 
If PBHs are clustered inside our own Galactic halo, then there should also be a Galactic $\gamma$-ray background and, since
this would be anisotropic, it  should be separable from the extragalactic background. The ratio of the anisotropic to
isotropic intensity depends on the Galactic longtitude and latitude, the Galactic core radius and the halo flattening. Wright claims that such a halo background has been detected \cite{w96}. His detailed fit to the EGRET data, subtracting various other known components, requires the PBH clustering factor to be $(2-12) \times 10^5 h^{-1}$, comparable to that expected.

\subsection{Antiprotons and Antideuterons}

\begin{figure}[htbp]
\begin{center}
\includegraphics[scale=0.4]{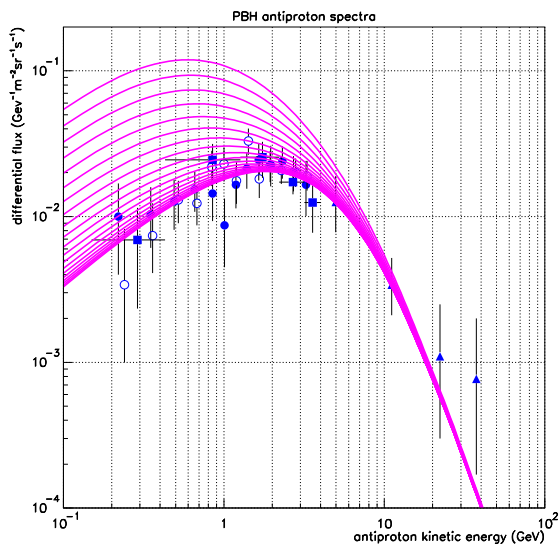}
\caption{Comparison of PBH emission and antiproton data from
Barrau et al.}
\label{fig:bc7}
\end{center}
\end{figure}

Since the ratio of antiprotons to protons in cosmic rays is less than
$10^{-4}$ over the energy range $100~\mbox{MeV} - 
10~\mbox{GeV}$,
whereas PBHs should produce them in equal numbers, PBHs could only
contribute appreciably to the antiprotons \cite{t82}. It is usually assumed 
that the observed
antiproton cosmic rays are secondary particles, produced by 
spallation
of the interstellar medium by primary cosmic rays. However, 
the spectrum of secondary antiprotons should show a steep
cut-off at kinetic energies below 2 GeV, whereas the spectrum of PBH antiprotons should increase with decreasing energy down to 0.2 GeV. Also the antiproton fraction should tend to 0.5 at low energies, so these features provide
a distinctive signature \cite{kir}. 
MacGibbon \& Carr originally calculated the PBH density required to explain the interstellar antiproton flux at 
1 GeV and found a value somewhat larger than the $\gamma$-ray limit \cite{mc}. More recent data on the antiproton flux below 0.5 GeV come from the BESS balloon experiment \cite{yosh} and Maki et al. \cite{mmo} tried to fit this in the PBH scenario by 
using Monte Carlo simulations of cosmic ray propagation. 
A more recent attempt to fit the antiproton data comes from Barrau et al. \cite{barrau} and is shown in Fig.\ref{fig:bc7} PBHs might also be detected by their antideuteron flux \cite{barrauetal}.

\subsection{PBH Explosions}
One of the most striking observational consequences of PBH evaporations would be their final explosive phase. However, 
in the standard particle physics picture, where the number of elementary particle species never exceeds around 100, the
likelihood of detecting such explosions is very low. Indeed, in 
this case, observations only place an upper limit
on the explosion rate of $ 5 \times10^8 {\rm pc}^{-3}{\rm y}^{-1}$ \cite{alex,sem}. This 
compares to Wright's $\gamma$-ray halo limit of 0.3 ${\rm pc}^{-3}{\rm y}^{-1}$ \cite{w96} and the Maki et al. antiproton limit of 0.02 ${\rm pc}^{-3}{\rm y}^{-1}$ \cite{mmo}.

However, the physics at the QCD phase transition is
still uncertain and the prospects of detecting explosions would
be improved in less conventional particle physics models. For example, in a Hagedorn-type picture, where the number of particle species exponentiates at the the quark-hadron temperature, 
the upper limit is reduced to 0.05 ${\rm pc}^{-3}{\rm y}^{-1}$ \cite{fic}.
Cline and colleagues have argued that one might
expect the formation of a QCD fireball at this temperature \cite{cho,csh} and this might even explain some of the short period $\gamma$-ray bursts observed by BATSE.
They claim to have found 42 candidates of this kind and the fact that their distribution matches the spiral arms suggests that they are Galactic \cite{cmo}. Although this proposal is speculative and has been disputed \cite{g01}, it has the attraction of making testable predictions (eg. the
hardness ratio of the burst should increase as the duration decreases). A rather different way of producing a $\gamma$-ray burst is to assume that the outgoing charged particles form a plasma due to turbulent magnetic 
field effects at sufficiently high temperatures \cite{bel}.

Some people have emphasized the possibility of detecting
very high energy cosmic rays from PBHs using air shower techniques \cite{hzmw,css,klw}. However, this is refuted by the claim of Heckler \cite{hec1} that QED interactions could
produce an optically thick photosphere once the black hole temperature exceeds $T_{crit}=45$ GeV. In this case, the mean photon
energy is reduced to $m_e(T_{BH}/T_{crit})^{1/2}$, which is well below $T_{BH}$, so
the number of high energy photons is much reduced. He
has proposed that a similar effect may operate at even lower temperatures due to QCD effects \cite{hec2}. Several groups have examined the implications of this proposal for PBH emission \cite{cms,kap}. However, these 
arguments should not be regarded as definitive since MacGibbon et al. claim that QED and QCD interactions are never important \cite{mcp}.

\section{PBHs as a Probe of Quantum Gravity}

\subsection{Planck Mass Relics}

Some people have speculated that black hole evaporation could cease once the hole gets close to the Planck mass \cite{mae,bow,cpw}. For example, in the standard Kaluza-Klein picture, extra dimensions are assumed to be compactified on the scale of the Planck length. This means that the influence of these extra dimensions becomes important at the energy scale of $10^{19}$GeV. Such effects could conceivably result in evaporation ceasing at the Planck mass. Various non-quantum-gravitational effects (such as higher order corrections to the gravitational Lagrangian or string effects) could also lead to stable relics \cite{cgl} but the relic mass is usually close to the Planck scale. 

Another possibility, as argued by Chen \& Adler \cite{chenad}, is that stable relics could arise if one invokes a ``generalized uncertainty principle''. This replaces the usual uncertainty principle with one of the form
\begin{equation}
\Delta x > \frac{\hbar}{\Delta p} + l_P^2\frac{\Delta p}{\hbar},
\end{equation}
where the second term is supposed to account for self-gravity effects. This means that the black hole temperature becomes
\begin{equation}
T_{BH} = {M c^2 \over 4\pi k} \left(1-\sqrt{1-\frac{M_P^2}{M^2}} \right).
\end{equation}
This reduces to the standard Hawking form for $M \gg M_P$ but it remains finite instead of diverging at the Planck mass itself. 
Cavaglia and colleagues have extended this argument to higher dimensional black holes \cite{cav} and also included the effects of thermal fluctuations  \cite{cd}. 

Whatever the cause of their stability, Planck mass relics would provide a possible cold dark matter candidate \cite{m87}. Indeed this leads to the ``relics" constraints indicated in Fig.1, Fig.2 and Fig.3.  In particular, such relics could derive from inflationary PBHs \cite{bcl,bkp}. If the relics have a mass $\kappa M_P$, then the requirement that they have less than the critical density implies \cite{cgl}
\begin{equation}
\beta(M) < 10^{-27}\kappa^{-1} (M/M_P)^{3/2} 
\end{equation}
for the mass range
\begin{equation}
(T_{RH}/T_{P})^{-2}M_{Pl} < M < 10^{11}\kappa^{2/5}M_P. 
\end{equation}
The upper mass limit arises because PBHs larger than this dominate the total density before they evaporate.  Producing a critical density of relics  obviously requires fine-tuning of the index $n$. Also one needs $n\approx$1.3, which is barely compatible with the WMAP results \cite{ask,bbbp}.  Nevertheless, this is possible in principle. In particular, it has been argued that hybrid inflation could produce relics from $10^5$g PBHs formed at $10^{-32}$s  \cite{chen,thom}.  

\subsection{PBHs and Brane Cosmology}

In ``brane" versions of Kaluza-Klein theory, some of the extra dimensions can be much larger than the usual Planck length and this means that quantum gravity effects may become important at a much smaller energy scale than usual.
If the internal space has $n$ dimensions and a compact volume $V_n$,  then Newton's constant $G_N$ is related to the higher dimensional gravitational constant $G_D$ and the value of the modified Planck mass $M_P$ is related to the usual 4-dimensional Planck mass $M_4$ by the order-of-magnitude equations:
\begin{equation}
G_N \sim G_D/V_n, \quad M_P^{n+2} \sim M_4^2/V_n.
\end{equation}
The same relationship applies if one has an infinite extra dimension but with a ``warped" geometry, provided one interprets $V_n$ as the ``warped volume". In the standard model, $V_n \sim 1/M_4^n$ and so $M_P \sim M_4$. However, with large extra dimensions, one has $V_n \gg 1/M_4^n$ and so $M_P \ll M_4$. 

Brane cosmology modifies the standard PBH formation and evaporation scenario. In particular, in Randall-Sundrum type II scenarios, a $\rho ^2$ term appears in the Friedmann equation and this dominates the usual $\rho$ term for
\begin{equation}
T > 10^{18} (l/l_P)^{-1/4} {\rm GeV} 
\end{equation}
where $l$ is the scale of the extra dimension.  This exceeds 1~TeV for $ l < 0.2$ mm. Black holes which are smaller than $l$ are effectively 5-dimensional. This means that they are cooler and evaporate more slowly than in the standard scenario:
\begin{equation}
T_{BH} \propto M^{-1/2}, \quad \tau \propto M^2.
\end{equation}
The critical mass for PBHs evaporating at the present epoch is now in the range $3\times10^{9}$g to $4\times10^{14}$g. This modifies the standard PBH evaporation constraints \cite {gcl,cgl2} and, if some cosmic rays come from PBHs, it means that these could probe the extra dimensions \cite{sns,send}. Another complication is that accretion could dominate evaporation during the era when the $\rho^2$ term dominates the $\rho$ term in the Friedmann equation \cite{maj}.

\subsection{Black Hole Production at Accelerators}

One exciting implication of these scenarios is that quantum gravitational effects may arise at the experimentally observable TeV scale.
If so, this would have profound implications for black hole formation and evaporation since black holes could be generated in accelerator experiments, such as the Large Hadron Collider (LHC) \cite{dl,gt,rt}. Although the black holes produced in accelerators should not themselves be described as ``primordial", since they do not form in the early Universe, it is clear that these experiments will also have profound implications for PBHs because, at sufficiently early times, the effects of the extra dimensions will be cosmologically important (see Section 7.2).

Two partons with centre-of-mass energy $\sqrt{s}$ will form a black hole if they come within a distance corresponding to the Schwarzschild radius $r_S$ for a black hole whose mass $M_{BH}$ is equivalent to that energy \cite{gt}. Thus the cross-section for black hole production is
\begin{equation}
\sigma_{BH} \approx \pi r_S^2 \Theta(\sqrt{s}-M_{BH}^{min})
\end{equation}
where $M_{BH}^{min}$ is the mass below which the semi-classical approximation fails. Here the Schwarzschild radius itself depends upon the number of internal dimensions:
\begin{equation}
r_S \approx {1\over M_P}\left({M_{BH}\over M_P}\right)^{1/(1+n)},
\end{equation}
so that $\sigma_{BH} \propto s^{2/(n+1)}$. This means that the cross-section for black hole production in scattering experiments goes well above the cross-section for the standard model above a certain energy scale and in a way which depends on the number of extra dimensions.

The evaporation of the black holes produced in this way will produce a characteristic signature \cite{dl,gt,rt} because the temperature and lifetime of the black holes depend on the number of internal dimensions:
\begin{equation}
T_{BH} \sim {n+1\over r_S}, \quad \tau_{BH} \sim {1\over M_P}\left({M_{BH}\over M_P}\right)^{(n+3)/(n+1)}.
\end{equation}
Thus the temperature is decreased relative to the standard 4-dimensional case and the lifetime is increased. The important qualitative effect is that a large fraction of the beam energy is converted into transverse energy, leading to large-multiplicity events with many more hard jets and leptons than would otherwise be expected. In principle, the formation and evaporation of black holes might be observed by the LHC when it turns on in 2007 and this might also allow one to experimentally probe the number of extra dimensions. On the other hand, this would also mean that scattering processes above the Planck scale could not be probed directly because they would be hidden behind a black hole event horizon \cite{gid}.  

Similar effects could be evident in the interaction between high energy cosmic rays and atmospheric nucleons. Nearly horizontal cosmic ray neutrinos would lead to the production of black holes, whose decays could generate deeply penetrating showers  with an electromagnetic component substantially larger than that expected with conventional neutrino interactions. Several authors have studied this in the context of the Pierre Auger experiment, with event rates in excess of one per year being predicted \cite{ag,fs,rt}. Indeed there is a small window of opportunity in which Auger might detect such events before the LHC. On the other hand, it must cautioned that these estimates may be overly optimistic. Others have claimed that black hole production in accelerators may be unobservable even if TeV quantum gravity is correct \cite{ahn}. 

\section{Conclusions}
We have seen that PBHs could be used to study the early Universe, gravitational collapse, high energy physics and quantum gravity. In the ``early Universe" context, useful constraints can be placed on inflationary scenarios. In the ``gravitational collapse" context, the existence of PBHs could provide a test of critical phenomena. In the ``high energy physics" context, information may come from cosmic ray and even gamma-ray bursts if suitable physics is invoked at the QCD phase transition. In the ``quantum gravity" context, the formation and evaporation of small black holes could be observable in cosmic ray events and accelerator experiments if the quantum gravity scale is around a TeV.

\end{document}